\documentclass[aps, prb, twocolumn, floatfix, amsfonts, amsmath, amssymb, superscriptaddress, letterpaper, 10pt]{revtex4-1}

\usepackage{hyperref}
\usepackage{cleveref}
\usepackage{color, graphicx, physics, ulem}

\newcommand{\req}[1]{Eq.~(\ref{#1})}
\newcommand{\reqs}[1]{Eqs.~(\ref{#1})}
\newcommand{\rref}[1]{(\ref{#1})}

\newcommand{\ee}{\mathrm{e}}
\newcommand{\ii}{\mathrm{i}}
\newcommand{\ns}{\qty(\vb{n} \cdot \hat{\vec{\sigma}})}
\newcommand{\doubargs}{\qty(\vb{x}, \vb{x}'; \omega)}

\newcommand{\rad}{\sqrt{\abs{\Delta}^{2} - \omega^{2}}}
\newcommand{\solid}{\Omega_{\hat{\vb{k}}}}
\newcommand{\scale}{0.9}

\begin{document}
\title{Odd-frequency superconductivity near a magnetic impurity in a conventional superconductor}
\author{Dushko Kuzmanovski}
\affiliation{NORDITA, Stockholm University, Roslagstullsbacken 23, SE-106 91 Stockholm, Sweden}
\author{Rub\'{e}n Seoane Souto}
\affiliation{Division of Solid State Physics and NanoLund,
Lund University, Box 118, S-22100 Lund, Sweden}
\affiliation{Center for Quantum Devices and Station Q Copenhagen, Niels Bohr Institute,
University of Copenhagen, DK-2100 Copenhagen, Denmark}
\author{Alexander~V.~Balatsky}
\affiliation{NORDITA, Stockholm University, Roslagstullsbacken 23, SE-106 91 Stockholm, Sweden}
\affiliation{Department of Physics, University of Connecticut, Storrs, CT 06269, USA}
\date{\today}

\begin{abstract}
Superconductor-ferromagnetic heterostructures have been suggested as one of the most promising alternatives of realizing odd-frequency superconductivity. In this work we consider the limit of shrinking the ferromagnetic region to the limit of a single impurity embedded in a conventional superconductor, which gives raise to localized Yu-Shiba-Rusinov (YSR) bound states with energies inside the superconducting gap. We demonstrate that all the sufficient ingredients for generating odd-frequency pairing are present at the vicinity of these impurities. We investigate the appearance of all possible pair amplitudes in accordance with the Berezinskii $SP^{\ast}OT^{\ast} = -1$ rule, being the symmetry under the exchange of spin, spatial, orbital (in our case $O=+1$) and time index, respectively. We study the spatial and frequency dependence of of the possible pairing amplitudes, analyzing their evolution with impurity strength and identifying a reciprocity between different symmetries related through impurity scattering. We show that the odd-frequency spin-triplet pairing amplitude dominates at the critical impurity strength, where the YSR states merge at the middle of the gap, while the even components cancel out close to the impurity. We also show that the spin-polarized local density of states exhibits the same spatial and frequency behavior as the odd-$\omega$ spin-triplet component at the critical impurity strength.
\end{abstract}
\maketitle

\section{\label{sec:Intro}Introduction}
Odd-frequency (odd-$\omega$) superconducting (SC) pairing is a proposed unconventional dynamic SC state that is both non-local and odd in the relative time coordinate.~\cite{Berezinskii74} Berezinskii was the first to point out that the only requirement on the pair correlator is to be antisymmetric under a simultaneous exchange of spin ($S$), spatial ($P^{\ast}$) and time ($T^{\ast}$) labels,~\cite{Berezinskii74, AVB92} which was later extended for multiorbital (multiband) systems~\cite{AMBSMultiOrb2013} to include exchange of the orbital index ($O$). It is written concisely as $S P^{\ast} O T^{\ast} = -1$ and referred to as Berezinskii rule. By allowing for odd time (or, equivalently, frequency) dependence, the two possible symmetries (spin-singlet even-parity and spin-triplet odd-parity) were extended to two more (spin-singlet odd-parity and spin-triplet even-parity).~\cite{JL_AVB_OddwSC17} The odd-$\omega$ spin-triplet and even-parity order was initially proposed  in liquid ${}^{3}\mathrm{He}$ and referred to as Berzinskii state. It was eventually ruled out in favor of  the even-frequency, spin-triplet, odd-parity.~\cite{Leggett1975, Wheatley1975} This idea drove the interest on realizing odd-$\omega$ states in solids. Later, it has been shown that phonon-mediated electron-electron interactions cannot stabilize an odd-$\omega$ SC order parameter (OP),~\cite{Abrahams93}, but other mechanisms based on spin-fluctuation-mediated interactions dependent on the Cooper pairs spin may be favorable. Several systems, such as disordered systems,~\cite{Kirkpatrick91, Belitz92, BelitzAndMott94} heavy fermion and Kondo systems,~\cite{Coleman93, Coleman94, Coleman95, CoxKondoEff98} and, more recently, Dirac semimetals~\cite{SukhachovOddwDirac19} were contemplated as theoretical possibilities of stabilizing an odd-$\omega$ SC OP. These materials were shown to exhibit exotic electromagnetic properties.~\cite{AbrahamsOddwProp95, SukhachovResponse19}
 
A different approach to inducing odd-$\omega$ pair correlations is through engineering heterostructures where a conventional (even-frequency, spin-singlet, $s$-wave) SC is proximitized to region that causes the breaking of some of its symmetries. This generates a corresponding odd-$\omega$ correlation component (see Sec.~I.~C. in Ref.~\onlinecite{JL_AVB_OddwSC17}). Historically  the first and the most analyzed systems are the superconductor-ferromagnet (SF) heterojunctions where the odd-$\omega$ spin-triplet $s$-wave pair correlations survive disorder.~\cite{Bergeret2001a, Kadigrobov2001} Issues concerning various geometries, magnetization profiles and the effects on different proximity and inverse proximity induced orders were considered (for a recent review see~\cite{BergeretSF05}). More recently, the interest in these heterojunctions was revived in relation to their potential application in superconducting spintronic devices.~\cite{JLSCSpintronics15, EschrigSpinSCCurr15, DiBernardoSFjunc15, HolmqvistSpinTransport18}

\begin{figure*}
\includegraphics[width=0.95 \linewidth]{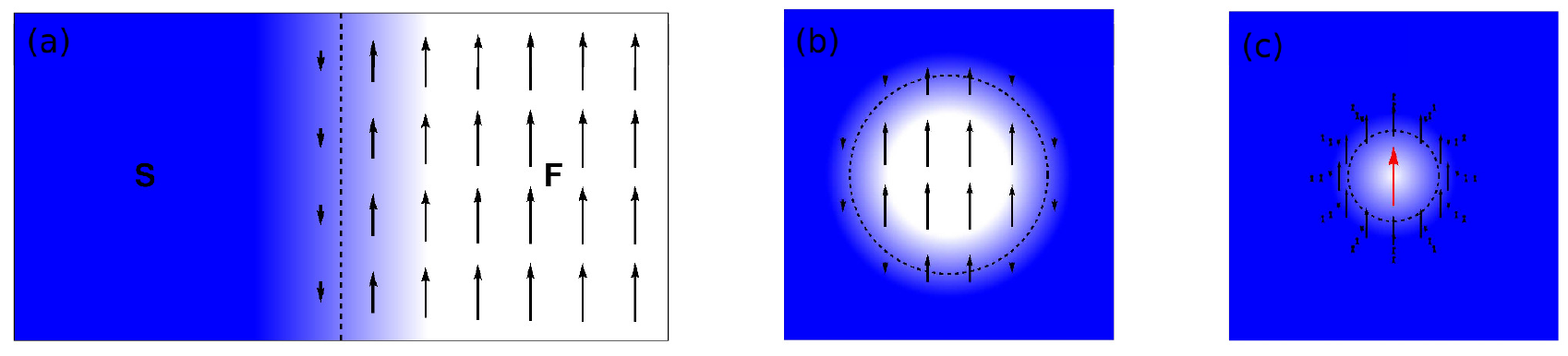}
\caption{\label{fig:Concept}The conceptual decrease of the size of the magnetic region: (a) an SF heterojunction. The blue colors saturation represents the magnitude of the conventional pair correlations, encoding the proximity effect in the ferromagnetic (F) region, and the inverse proximity effect in the S region. The arrow length is the magnitude of the local magnetization, leaking in the S region; (b) a ferromagnetic island inside a bulk SC; (c) a single magnetic impurity immersed in a bulk SC. The red arrow is the fixed magnetic moment of the impurity.}
\end{figure*}

Conceptually, the sufficient ingredients for generating odd-$\omega$ pair correlations by proximity effect may be preserved while  transitioning from the geometry of heterostrutures, through finite-sized ferromagnetic islands to the limit of an isolated magnetic impurity, as illustrated in Fig.~\ref{fig:Concept}. In this work we consider the extreme case of reducing the size of the ferromagnetic island, which corresponds to the controlled immersion of a single magnetic impurity atom in a clean superconductor, panel (c) in Fig.~\ref{fig:Concept}.

These magnetic impurities were instrumental in mapping out the spatial symmetry of the order parameter through scanning tunneling spectroscopic measurements in the cuprate superconductors.~\cite{AVBImpStates06} A well-known result~\cite{Yu1965, Shiba1968, Rusinov1969} is the appearance of spin-polarized sub-gap Yu-Shiba-Rusinov (YSR) bound states localized around the impurity position, whose energy can cross zero as the impurity strength is varied. A disordered dense collection of such impurities causes a filling of the energy gap and suppression of bulk superconductivity. 
There has been a renewed experimental,~\cite{NadjPerge14, Ruby15, Pawlak16,Haonan_arXiv19} and theoretical \cite{Choy2011, NadjPerge2013, Klinovaja2013, Braunecker2013, Vazifeh2013, Pientka2013, Nakosai2013, Poeyhoenen2014, Kim2014, Reis2014, Brydon2015, Bjoernson15,  Roentynen2015, Li2016, Teixeira2019} interest in magnetic impurities on superconductors, mostly owing to the possibility of a realization of a one-dimensional topological superconductor that host zero-energy Majorana end modes in the presence of spin-orbit coupling.

In this work, we consider the limiting situation of magnetic defects (Fig.~\ref{fig:Concept}(c)), namely an isolated magnetic impurity atom immersed in a bulk conventional SC. It is explicitly demonstrated that odd-$\omega$ pair correlations are generated, in accordance with the $S P^{\ast} O T^{\ast}$ rule.Very recently, a complementary study has demonstrated the existence of odd-$\omega$ in non-magnetic potential impurities due to the renormalization of the SC OP. This corresponds to the limiting case of a metal-superconductor (NS) heterojunctions (where the N region has been reduced to an impurity), shown to to host odd-$\omega$ pair correlations much later than the already mentioned cases of SF heterojunctions when the spatial modulation of the SC OP is taken into account.~\cite{Tanaka2007a, Tanaka2007b, Tanaka2007c}.

The simple geometry under consideration offers the advantage of exactly decomposing the spin, spatial and frequency dependence of the pair correlations, enabling a clear demonstration of the conversion of spin and time symmetry. Motivated by the experimental capabilities in measuring magnetic impurities in superconductors,~\cite{YazdaniSpectReview16, CornilsYSRSpec17} it is worthwhile to point out the features of the odd-$\omega$ Berezinskii pairing that are correlated with spectroscopic data.~\cite{Salkola1997, Kim2015, Pershoguba15, Bjoernson2017} We discuss the behavior of the spin-polarized (SP) local density of states (LDOS) as it can be related to the pair correlations. 

The rest of the article is organized as follows: In Sec.~\ref{sec:Model} we describe the model Hamiltonian of the system formed by a Kondo-type impurity immersed in a BCS bulk SC. We introduce the Green's function and the Dyson equation in Sec.~\ref{sec:DysonEqn}. The results are presented in Sec.~\ref{sec:Results}. In Sec.~\ref{sec:AnalyticGimp}, we provide analytical expressions for the impurity Green's function. In Sec.~\ref{sec:Reciprocity}, the reciprocity relations between the pair correlation components are introduced, discussing the parameters where odd-$\omega$ components dominate. In Sec.~\ref{sec:spLDOS} we show the SP LDOS  close to the impurity and local magnetization DOS. In Sec.~\ref{sec:ResLocalPairCorr}, \ref{sec:ResNonLocalPairCorr} the pair correlations are represented. Finally, analytic expressions and derivation details are given in the Appendices.

\section{\label{sec:Model}Model and Method}
We consider a clean metal with spin-degenerate and isotropic dispersion $\xi_{\vb{k}} \approx v_{\rm F} \, \qty(\abs{\vb{k}} - k_{\rm F})$ in $D$ spatial dimensions whose Fermi surface is a sphere of radius $k_{\rm F}$ and the band velocity at the Fermi level (Fermi velocity) is $v_{\rm F}$. The metal is an intrinsic conventional SC, with a spatially uniform~\footnote{The presence of the magnetic impurity alters the local equal-time pair correlator in its vicinity. Thus, a self-consistent solution for the SC OP would result in a non-uniform SC OP around the impurity. But, since we consider a single impurity immersed in a continuum capable of screening its effects, this non-uniformity is only a second-order correction and not of main concern for the purposes of this work.} singlet $s$-wave SC OP $\Delta$. The Hamiltonian of the system is given by $\mathcal{H} = \mathcal{H}_{\rm BCS} + \mathcal{H}_{\rm imp}$, where the metal is a BCS mean-field Hamiltonian:
\begin{equation}
\label{eq:BCSHam}
\mathcal{H}_{\rm BCS} = \sum_{\vb{k}\sigma} \xi_{\vb{k}} \, c^{\dagger}_{\vb{k}\sigma} \, c_{\vb{k}\sigma} + \sum_{\vb{k}} \Bqty{\Delta \, c^{\dagger}_{\vb{k}\uparrow} \, c^{\dagger}_{-\vb{k}\downarrow} + \rm{H. c.}}.
\end{equation}
Here, $c_{\vb{k}\sigma}$ ($c^{\dagger}_{\vb{k}\sigma}$) is the second-quantized electron annihilation (creation) operator that destroys (creates) an electron with wavevector $\vb{k}$ and spin projection $\sigma \in \Bqty{\uparrow, \downarrow}$.

Inside the metal, a point-like magnetic impurity at position $\vb{x} = \vb{0}$ is immersed. The impurity Hamiltonian is a Kondo-type Hamiltonian described by
\begin{equation}
\label{eq:KondoHam}
\mathcal{H}_{\rm imp} = U_{m} \, \qty(\vb{n} \cdot \vb{M}\qty(0)),
\end{equation}
where $U_{m}$ is the impurity strength and $\vb{n}$ is a unit vector ($\qty(\vb{n} \cdot \vb{n}) = 1$) that determines the polarization of the magnetic impurity. $\vb{M}\qty(\vb{x})$ is the real-space spin polarization due to the conduction electrons:
\begin{equation}
\label{eq:Magnetization}
M_{i}\qty(\vb{x}) = \sum_{\sigma \sigma'} \qty[\sigma_{i}]_{\sigma\sigma'} c^{\dagger}_{\sigma}\qty(\vb{x}) \, c_{\sigma'}\qty(\vb{x}),
\end{equation}
with $i = \qty(x, y, z)$ being a Cartesian component label. Here, $c^{\dagger}_{\sigma}\qty(\vb{x})$ is a creation operator that creates an electron at position $\vb{x}$ with spin polarization $\sigma$. It is related to the momentum-space via a Fourier transform:
\begin{equation}
\label{eq:RealKspaceCreation}
c^{\dagger}_{\sigma}\qty(\vb{x}) = \frac{1}{L^{D/2}} \, \sum_{\vb{k}} \ee^{-\ii \, \vb{k} \cdot \vb{x}} \, c^{\dagger}_{\vb{k}\sigma},
\end{equation}
and analogously for $c_{\sigma}\qty(\vb{x})$. The minus sign in the exponent is in accordance with the choice for the translation operator $T\qty(\vb{a}) = \exp\qty[-\ii \, \qty(\vb{a} \cdot \vb{P})]$ (we choose units in which $\hbar = k_{\rm B} = 1$) so that $T\qty(\vb{a}) \, c^{\dagger}_{\sigma}\qty(\vb{x}) \, T^{\dagger}\qty(\vb{a}) = c^{\dagger}_{\sigma}\qty(\vb{x} + \vb{a})$. The normalization $L^{-D/2}$ ensures that the real-space second-quantized operators satisfy the anticommutation relation:
\begin{equation}
\Bqty{c_{\sigma}\qty(\vb{x}), c^{\dagger}_{\sigma'}\qty(\vb{x}')}  = \delta_{\sigma\sigma'} \, \delta^{D}\qty(\vb{x} - \vb{x}').
\end{equation}
Thus, $\qty[c^{\dagger}_{\sigma}\qty(\vb{x})] = \qty[c_{\sigma}\qty(\vb{x})] = L^{-D/2}$, and $\qty[M] = L^{-D}$. In order that $\mathcal{H}_{\rm imp}$ has dimension of energy, $U_{m}$ ought to have dimension $\qty[U_{m}] = {\rm L}^{D} \, \qty[E]$.

We define the Nambu spinor:
\begin{equation}
\label{eq:NambuSpinor}
\Psi\qty(\vb{x}) = \mqty( c_{\sigma}\qty(\vb{x}) &,& \qty(\ii \, \sigma_{y})_{\sigma \sigma'} \, c^{\dagger}_{\sigma'}\qty(\vb{x}) )^{\top},
\end{equation}
which satisfies the commutation relations with the spin operator:
\begin{equation}
\label{eq:SpinRot}
-\qty[S_{a}, \Psi\qty(\vb{x})] = \frac{1}{2} \, \qty(\hat{\tau}_{0} \otimes \hat{\sigma}_{a}) \cdot \Psi\qty(\vb{x}),
\end{equation}
where the Pauli matrices $\hat{\tau}_{\mu}$ and $\hat{\sigma}_{\mu}$ operate in the Nambu and the spin sub-spaces, respectively. We define an imaginary-time-ordered Green's function matrix:
\begin{eqnarray}
& \check{G}\qty(\vb{x}, \tau; \vb{x}', \tau') = -\expval{T_{\tau} \, \Psi\qty(\vb{x}, \tau) \otimes \bar{\Psi}\qty(\vb{x}', \tau')} \nonumber \\
& = \left(\begin{array}{c|c}
\hat{G}^{e e}\qty(\vb{x}, \tau; \vb{x}', \tau') & \hat{G}^{e h}\qty(\vb{x}, \tau; \vb{x}', \tau') \\
\hline
\hat{G}^{h e}\qty(\vb{x}, \tau; \vb{x}', \tau') & \hat{G}^{h h}\qty(\vb{x}, \tau; \vb{x}', \tau')
\end{array}\right), \label{eq:GreenNambuSpinTime}
\end{eqnarray}
where $\check{}$ denotes a $4\times4$ matrix in the Nambu-spin space. The imaginary-time Heisenberg operators evolve with the total Hamiltonian of the system $\mathcal{H}$ according to:
\begin{subequations}
\label{eq:ImagTimeEvol}
\begin{eqnarray}
& \Psi\qty(\vb{x}, \tau) = \ee^{\tau \, \mathcal{H}} \, \Psi\qty(\vb{x}) \, \ee^{-\tau \, \mathcal{H}}, \label{eq:PsiTimeEvol} \\
& \bar{\Psi}\qty(\vb{x}, \tau) = \ee^{\tau \, \mathcal{H}} \, \Psi^{\dagger}\qty(\vb{x}) \, \ee^{-\tau \, \mathcal{H}}. \label{eq:PsiBarTimeEvol}
\end{eqnarray}
\end{subequations}
In case of a time-independent Hamiltonian, the Green's function \req{eq:GreenNambuSpinTime} is only a function of the time difference $\tau - \tau'$ and may be expanded in fermionic Matsubara frequencies:
\begin{equation}
\label{eq:GreenNambuSpinFreq}
\check{G}\qty(\vb{x}, \tau; \vb{x}', \tau') = \frac{1}{\beta} \, \sum_{\omega_{m}} \ee^{-\ii \, \omega_{m} \, \qty(\tau - \tau')} \, \check{G}\qty(\vb{x}, \vb{x}'; \ii \, \omega_{m}).
\end{equation}
Then, through an analytic continuation $\ii \, \omega_{m} \rightarrow \omega$, the Green's function becomes a function of complex frequency.

In the matrix structure adopted in Eq. \req{eq:NambuSpinor}, the $SP^{\ast}OT^{\ast} = -1$ rule has the form:
\begin{subequations}
\begin{equation}
\label{eq:SPOTinTime}
\hat{G}^{e h}\qty(\vb{x}, \tau; \vb{x}', \tau') = + \qty[ \hat{\sigma}_{y} \cdot \hat{G}^{e h}\qty(\vb{x}', \tau'; \vb{x}, \tau) \cdot \hat{\sigma}_{y}]^{\top},
\end{equation}
or, equivalently, as a function of Matsubara frequency:~\cite{Bzudshek2015}
\begin{equation}
\label{eq:SPOTinFreq}
\hat{G}^{e h}\qty(\vb{x}, \vb{x}'; \ii \, \omega_{m}) = +\qty[ \hat{\sigma}_{y} \cdot \hat{G}^{e h}\qty(\vb{x}', \vb{x}; -\ii \, \omega_{m}) \cdot \hat{\sigma}_{y}]^{\top}.
\end{equation}
Therefore, \req{eq:SPOTinFreq} implies for the singlet $\qty[G^{e h}]^{s}$, and triplet components $\qty[\vb{G}^{e h}]^{t}$:
\begin{eqnarray}
& \hat{G}^{e h}\qty(\vb{x}, \vb{x}'; \ii \, \omega_{m}) = [G^{e h}]^{s}\qty(\vb{x}, \vb{x}'; \ii \, \omega_{m}) \, \hat{\sigma}_{0} \nonumber \\
& + \qty([\vb{G}^{e h}]^{t}\qty(\vb{x}, \vb{x}'; \ii \, \omega_{m}) \cdot \hat{\vec{\sigma}}), \label{eq:SpinDecomp}
\end{eqnarray}
the following symmetry properties under $P^{\ast}$ and $T^{\ast}$:
\begin{eqnarray}
& \qty[G^{e h}]^{s}\qty(\vb{x}, \vb{x}'; \ii \, \omega_{m}) = +\qty[G^{e h}]^{s}\qty(\vb{x}', \vb{x}; -\ii \, \omega_{m}), \nonumber \\
&P^{\ast} \, T^{\ast} = +1, \label{eq:PTSymmSinglet} \\
& \qty[\vb{G}^{e h}]^{t}\qty(\vb{x}, \vb{x}'; \ii \, \omega_{m}) = -\qty[\vb{G}^{e h}]^{t}\qty(\vb{x}', \vb{x}; -\ii \, \omega_{m}), \nonumber \\
&P^{\ast} \, T^{\ast} = -1. \label{eq:PTSymmTriplet}
\end{eqnarray}
This in accordance with the fact that spin-singlet pair correlations are odd under exchange of spin indices ($S = -1$), while spin-triplet are even ($S = +1$). 
\end{subequations}

\subsection{\label{sec:DysonEqn}Dyson equation}
The Green's function defined in \req{eq:GreenNambuSpinFreq} in the presence of impurity potential \req{eq:KondoHam} satisfies the following Dyson equation:
\begin{subequations}
\begin{eqnarray}
& \check{G}\doubargs = \check{G}_{\rm cl}\doubargs \nonumber \\
& + U_{m} \, \check{G}_{\rm cl}\qty(\vb{x}, \vb{0}; \omega) \cdot \check{P} \cdot \check{G}\qty(\vb{0}, \vb{x}'; \omega), \label{eq:DysonRealSpace}
\end{eqnarray}
\end{subequations}
with
\begin{equation}
    \check{P} = \check{P}^{\dagger} = \check{P}^{-1} = \hat{\tau}_{0} \otimes \ns. \label{eq:ImpVertex}
\end{equation}
Here, $\check{G}_{\rm cl}\doubargs$ is the Green's function matrix for the clean system which is invariant under spatial translations:
\begin{equation}
\label{eq:GreenClean1}
\check{G}_{\rm cl}\doubargs = \frac{1}{L^{D}} \, \sum_{\vb{k}} \check{G}_{\rm cl}\qty(\vb{k}; \omega) \, \ee^{\ii \, \vb{k} \cdot \qty(\vb{x} - \vb{x}')},
\end{equation}
and $\check{G}_{\rm cl}\qty(\vb{k}; \omega)$ is assumed to correspond to a conventional $s$-wave spin-singlet superconductor with a SC OP $\Delta$ and an isotropic normal dispersion $\xi_{k}$ invariant under spin rotations. This means that $\check{G}_{\rm cl}\qty(\xi_{k}; \omega) = \hat{G}_{\rm cl}\qty(\xi_{k}; \omega) \otimes \hat{\sigma}_{0}$, where:
\begin{subequations}
\begin{eqnarray}
& \hat{G}^{-1}_{\rm cl}\qty(\xi_{k}; \omega) = \omega \, \hat{\tau}_{0} - \xi_{k} \, \hat{\tau}_{3} - \hat{\Delta}, \label{eq:BdGEqn1} \\
& \hat{\Delta} = \Delta \, \hat{\tau}_{+} + \Delta^{\ast} \, \hat{\tau}_{-}. \label{eq:SCvertex}
\end{eqnarray}
\end{subequations}
The integral over $\vb{k}$ in \req{eq:GreenClean1} is evaluated in the wide band limit in Appendix~\ref{app:RealSpaceGcl} and the result is expressible in terms of the functions $\qty(\phi_{c/s})_{D}\qty(r, \omega)$. Due to the translation invariance and isotropy, these functions only depend on $r = \abs{\vb{x} - \vb{x}'}$. The $\omega$-dependence enters through $\kappa\qty(\omega) = \rad/v_{\rm F}$, so their real part is even in $\omega$, while the imaginary part is odd. The real-space Green's function is given by:
\begin{subequations}
\label{eq:GreenClean2}
\begin{eqnarray}
& -\frac{1}{\pi \, N_{\rm F}} \, \hat{G}_{\rm cl}\doubargs \nonumber \\
& = \qty(\phi_{c})_{D}\qty(r; \omega) \, \hat{g}_{0}\qty(\omega) + \qty(\phi_{s})_{D}\qty(r; \omega) \, \hat{\tau}_{3}, \label{eq:GreenCleanSemiClassNambu} \\
& \hat{g}_{0}\qty(\omega) = \frac{\omega \, \hat{\tau}_{0} + \hat{\Delta}}{\rad}, \label{eq:Semi}
\end{eqnarray}
where $N_{\rm F}$ is the DOS per spin component and its dependence on the model parameters is given by \req{eq:NormDos}. Note that $\qty(\phi_{c})_{D}\qty(0; \omega) = 1$, and $\qty(\phi_{s})_{D}\qty(0; \omega) = 0$.
\end{subequations}
Finally, the solution of \req{eq:DysonRealSpace} is of the form:
\begin{equation}
\check{G}\doubargs = \check{G}_{\rm cl}\doubargs + \check{G}_{\rm imp}\doubargs, \label{eq:ImpGreen1}
\end{equation}
so that all the effects of the magnetic impurity potential are expressed in terms of the impurity Green's function $\check{G}_{\rm imp}\doubargs$. This function lacks translation symmetry due to the presence of a scattering center. Nevertheless, it has an analytic expression that may be obtained within the $T$-matrix approach. Details of its evaluation are given in Appendix~\ref{app:TmatExprs}. The $T$-matrix has a denominator that is a quadratic function of $\omega$, giving two poles at $\pm \varepsilon_{0}$, corresponding to the energy of the YSR states. They are given by:~\cite{AVBImpStates06}
\begin{equation}
\label{eq:YSRenergy}
\varepsilon_{0} = \frac{1 - J^{2}_{m}}{1 + J^{2}_{m}} \, \abs{\Delta},
\end{equation}
with the dimensionless impurity strength:
\begin{equation}
\label{eq:ImpStrength}
J_{m} = \pi \, N_{\rm F} \, U_{m}.
\end{equation}

\section{\label{sec:Results}Results}
In this section, we present results that are directly related to the pair correlations around the impurity and experimentally measurable single-particle properties. 

\begin{widetext}
\subsection{\label{sec:AnalyticGimp}Analytic expression for the impurity Green's function}
We begin our analysis deriving the expression for the impurity Green's function using the T-matrix formalism, described Sec. \ref{app:TmatExprs}. From \reqs{eq:ImpGreen2}, \rref{eq:TmatrixNambuSpin} and \rref{eq:GreenCleanSemiClassNambu}, we have the following explicit expression for the impurity correction to the Green's function matrix:
\begin{equation}
\label{eq:ImpGreen3}
-\frac{1}{\pi \, N_{\rm F}} \, \check{G}_{\rm imp}\doubargs = \frac{J_{m}}{D_{\rm YSR}\qty(\omega)} \, \qty[\hat{g}^{s}\doubargs \otimes \hat{\sigma}_{0} + \hat{g}^{t}\doubargs \otimes \ns],
\end{equation}
where
\begin{equation}
    \label{eq:DYSR}
    D_{\rm YSR}\qty(\omega) = \qty(1 + J^{2}_{m})^{2} \, \omega^{2} - \qty(1 - J^{2}_{m})^{2} \, \abs{\Delta}^{2}
\end{equation}
is a denominator with poles at the YSR bound-state energies $\pm\varepsilon_{0}$, \req{eq:YSRenergy}, and $\hat{g}^{s}\doubargs$ and $\hat{g}^{t}\doubargs$ denote the singlet and triplet component of the Green function. We note that the triplet component is aligned with the magnetization direction of the impurity. The singlet and triplet contributions can be decomposed in a basis of $4$ functions with definite parity under the exchange $\vb{x} \leftrightarrow \vb{x}'$:
\begin{subequations}
\label{eq:BasisFns}
\begin{eqnarray}
& \psi_{1}\qty(\vb{x}, \vb{x}'; \omega) = \qty(\phi_{c})_D\qty(r, \omega) \, \qty(\phi_{c})_D\qty(r', \omega) +\qty(\phi_{s})_D\qty(r, \omega) \, \qty(\phi_{s})_D\qty(r', \omega), \label{eq:Fn1} \\
& \psi_{2}\qty(\vb{x}, \vb{x}'; \omega) = \qty(\phi_{c})_D\qty(r, \omega) \, \qty(\phi_{c})_D\qty(r', \omega) -\qty(\phi_{s})_D\qty(r, \omega) \, \qty(\phi_{s})_D\qty(r', \omega), \label{eq:Fn2} \\
& \psi_{3}\qty(\vb{x}, \vb{x}'; \omega) = \qty(\phi_{s})_D\qty(r, \omega) \, \qty(\phi_{c})_D\qty(r', \omega) + \qty(\phi_{s})_D\qty(r', \omega) \, \qty(\phi_{c})_D\qty(r, \omega), \label{eq:Fn3} \\
& \psi_{4}\qty(\vb{x}, \vb{x}'; \omega) = \qty(\phi_{s})_D\qty(r, \omega) \, \qty(\phi_{c})_D\qty(r', \omega) - \qty(\phi_{s})_D\qty(r', \omega) \, \qty(\phi_{c})_D\qty(r, \omega). \label{eq:Fn4}
\end{eqnarray}
\end{subequations}
The first $3$ functions are even, while the last is odd under coordinate exchange. All of the functions, according to \reqs{eq:AngleInt}, are even in frequency. Analytic expressions and the spatial behavior of these functions are investigated in more detail in Appendix ~\ref{app:BasisFns} for different $D$ dimensions.
 
Using these basis functions, we may write $\hat{g}^{\alpha}\doubargs = \sum_{a = 1}^{4} \hat{g}^{\alpha}_{a}\qty(\omega) \, \psi_{a}\doubargs$, $\alpha \in \qty(s, t)$, where:

\begin{subequations}
\label{eq:NambuGreens}
\begin{eqnarray}
& \frac{\hat{g}^{t}_{1}\qty(\omega)}{1 - J^{2}_{m}} = -\frac{\rad \, \hat{g}^{s}_{1}\qty(\omega)}{2 J_{m} \, \omega} = \abs{\Delta}^{2} \, \hat{\tau}_{0} + \omega \, \hat{\Delta}, \label{eq:g1} \\
& \frac{\hat{g}^{t}_{2}\qty(\omega)}{\qty(1 + J^{2}_{m}) \, \omega \, \rad} = -\frac{\hat{g}^{s}_{2}\qty(\omega)}{J_{m} \, \qty[\qty(1 + J^{2}_{m}) \, \omega^{2} + \qty(1 - J^{2}_{m}) \, \abs{\Delta}^{2}]} = \hat{g}_{0}\qty(\omega), \label{eq:g2} \\
& \frac{\hat{g}^{t}_{3}\qty(\omega)}{\qty(1 + J^{2}_{m}) \, \omega \, \rad} = -\frac{g^{s}_{3}\qty(\omega)}{J_{m} \, \qty[\qty(1 + J^{2}_{m}) \, \omega^{2} + \qty(1 - J^{2}_{m}) \, \abs{\Delta}^{2}]} = \hat{\tau}_{3}, \label{eq:g3} \\
& \frac{\hat{g}^{t}_{4}\qty(\omega)}{\qty(1 - J^{2}_{m}) \, \rad} = -\frac{\hat{g}^{s}_{4}\qty(\omega)}{2 J_{m} \, \omega} = \hat{\tau}_3\hat{\Delta}. \label{eq:g4}
\end{eqnarray}
\end{subequations}
\end{widetext}

\subsection{\label{sec:Reciprocity}Reciprocal relations between singlet and triplet pair correlations}
The $SP^{\ast}OT^{\ast} = -1$ condition for single-band systems, where $O = +1$, goes over to $SP^{\ast}T^{\ast} = -1$. Thus, there are $4$ possibilities of allowed combined symmetries $\qty(S, P^{\ast}, T^{\ast})$. The enumeration of these possibilities, together with their labeling and the corresponding contributions from \reqs{eq:GreenClean2} and \rref{eq:NambuGreens} are given in Table~\ref{tab:TableGeh}.
\begin{table}[!h]
\begin{ruledtabular}
\begin{tabular}{||ccc|cc||}
$S$ & $P^{\ast}$ & $T^{\ast}$ & Notation & Contribution \\
\hline
$-$ & $+$ & $+$ & $G^{e h}_{(-,+,+)}\doubargs$ & $\hat{g}_{0}\qty(\omega)$, $\hat{g}^{s}_{1}\qty(\omega)$, $\hat{g}^{s}_{2}\qty(\omega)$ \\
$+$ & $+$ & $-$ & $G^{e h}_{(+,+,-)}\doubargs$ & $\hat{g}^{t}_{1}\qty(\omega)$, $\hat{g}^{t}_{2}\qty(\omega)$ \\
\hline
$+$ & $-$ & $+$ & $G^{e h}_{(+,-,+)}\doubargs$ & $\hat{g}^{t}_{4}\qty(\omega)$ \\
$-$ & $-$ & $-$ & $G^{e h}_{(-,-,-)}\doubargs$ & $\hat{g}^{s}_{4}\qty(\omega)$
\end{tabular}
\end{ruledtabular}
\caption{\label{tab:TableGeh}The allowed symmetries under $\qty(S, P^{\ast}, T^{\ast})$ for the electron-hole propagators $\hat{G}^{eh}\doubargs$, and the corresponding contributions. The different symmetries under $P^{\ast}$ are split, indicating the reciprocity in exchange of the symmetry under $S$ and $T^{\ast}$.}
\end{table}
The components $\qty[g^{t}_{a}]^{e h}\qty(\omega)$ for $a = 1, 2$ ($a = 3$ does not have an electron-hole components) are necessarily odd in $\omega$, while $\qty[g^{t}_{4}]^{e h}\qty(\omega)$ is even in $\omega$. For $\qty[g^{s}_{a}]^{e h}\qty(\omega)$, the situation is reversed. Having in mind the general matrix structure of $\hat{g}^{t}_{a}$ and $\hat{g}^{s}_{a}$ given in \reqs{eq:NambuGreens}, we see that for any pair of coordinates $\vb{x}$, $\vb{x}'$, there is a general reciprocity relation between the singlet and triplet components with different parity in frequency $\omega$. We point-out, however, that the basis functions \reqs{eq:BasisFns} also carry (even) frequency dependence, so, the total frequency dependence is coordinate dependent.

\begin{subequations}
Another reciprocity between the electron-hole components follows by noticing that the same ratio holds:
\begin{eqnarray}
& \hat{g}^{s}_{a}\qty(\omega) = r_{\rm I}\qty(\omega) \, \hat{g}^{t}_{a}\qty(\omega), \ (a = 1, 4), \nonumber \\
& r_{\rm I}\qty(\omega) = -\frac{2 J_{m} \, \omega}{\qty(1 - J^{2}_{m}) \, \rad}, \label{eq:Ratio1}
\end{eqnarray}
for two sets of basis functions with opposite parity under coordinate exchange. We note that at $\omega\to0$ the ratio vanishes, implying that $\hat{g}^{s}_{a}(\omega=0)=0$ ($a=1,4$). In the same way, at the critical impurity strength, $J_m=1$, $\hat{g}^{t}_{a}(\omega)=0$ ($a=1,4$). Taking the electron-hole components, we may rewrite the ratio in Eq. (\ref{eq:Ratio1}) as:
\begin{equation}
    \frac{\qty[g^{s}_{4}]^{e h}\qty(\omega)}{\qty[g^{s}_{1}]^{e h}\qty(\omega)} = \frac{\qty[g^{t}_{4}]^{e h}\qty(\omega)}{\qty[g^{t}_{1}]^{e h}\qty(\omega)} = \frac{\rad}{\omega}. \label{eq:Ratio1b}
\end{equation}
\end{subequations}
Thus, there is a reciprocity between exchange of the parity with respect to coordinate exchange ($a = 1$ and $a = 4$) and $\omega$, while keeping the same spin symmetry. The ratio of these pair correlations is independent of the impurity strength $J_{m}$.

A similar relation can be derived for the remaining two basis functions with same parity under coordinate exchange
\begin{eqnarray}
& \hat{g}^{s}_{a}\qty(\omega) = r_{\rm II}\qty(\omega) \, \hat{g}^{t}_{a}\qty(\omega), \ (a = 2, 3), \nonumber \\
& r_{\rm II}\qty(\omega) =  -\frac{J_{m} \, \qty[\qty(1 + J^{2}_{m}) \, \omega^{2} + \qty(1 - J^{2}_{m}) \, \abs{\Delta}^{2}]}{\qty(1 + J^{2}_{m}) \, \omega \, \rad}. \label{eq:Ratio2}
\end{eqnarray}
We note that $\qty[g^{\alpha}_{3}]^{eh}\qty(\omega)  = 0$, which may be interpreted as a selection rule, which guarantees that there is no electron-hole component generated by impurity scattering if neither the spin, nor the parity symmetry is changed.

\subsection{\label{sec:spLDOS}Spin-polarized local density of states}
We begin plotting local single-particle properties, such as the SP LDOS. The LDOS $\nu\qty(\vb{x}, \omega)$, and the LMDOS $\vb{m}\qty(\vb{x}, \omega)$ are evaluated as:
\begin{subequations}
\label{eq:Spectra}
\begin{eqnarray}
& \nu\qty(\vb{x}, \omega) = \nonumber \\
& -\frac{1}{\pi} \, \lim_{\eta \rightarrow 0} \Im\bqty{\Tr{\hat{G}^{ee}\qty(\vb{x}, \vb{x}, \omega + \ii \, \eta)}}, \label{eq:LDOS} \\
& \vb{m}\qty(\vb{x}, \omega) = \nonumber \\
& -\frac{1}{\pi} \, \lim_{\eta \rightarrow 0} \Im\bqty{\Tr{\hat{\vec{\sigma}} \cdot \hat{G}^{ee}\qty(\vb{x}, \vb{x}, \omega + \ii \, \eta)}}.
\end{eqnarray}
\end{subequations}
The magnetization has the same direction as the impurity spin polarization. The spin-up ($\sigma = +1$) and spin-down ($\sigma = -1$) components of the SP LDOS are evaluated as $\nu_{\sigma} = (\nu + \sigma \, \abs{\vb{m}})/2$, and can be experimentally measured with a magnetic tip STM in spectroscopic mode. From them, working backwards, one may determine $\vb{m}$.

When representing intensity plots of a quantity $f$ (SP LDOS, LMDOS, real or imaginary part of a particular pair correlation) containing narrow high-intensity peaks around the YSR bound-state energies or the gap edges, and having a smooth background, we use a scaling factor $s$ according to the sigmoid scaling function:
\begin{equation}
\label{eq:ScalingFn}
s\qty( f ) = \mathrm{sgn} \qty( f ) \, \qty[ 1 - \exp\qty(-\frac{\abs{f}}{f_{0}}) ].
\end{equation}
This is an odd-valued ($s\qty(-f) = -s\qty(f)$), smooth single-parameter function, where the parameter $f_{0}$ roughly corresponds to a region of values for $f$ for which the scaling is linear. This scaling has the advantage that it collapses the whole real line on a finite segment $\qty[-1, 1]$, while keeping a linear resolution for low magnitudes, which is particularly convenient in plots with narrow isolated peaks and a varying background.

\begin{figure}[!h]
\includegraphics[width=\scale \linewidth]{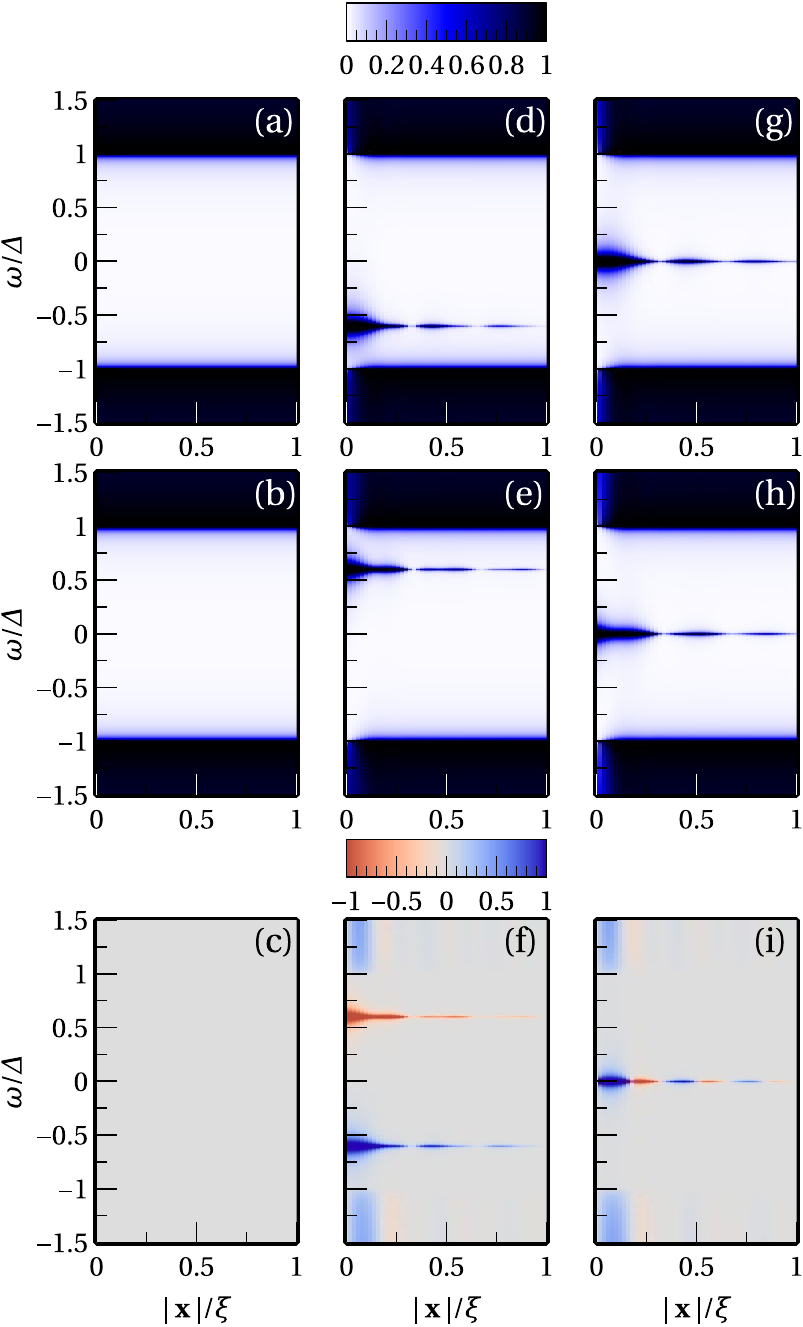}
\caption{\label{fig:LMDOS}The spin-up LDOS (top row, panels (a), (d), (g)), spin-down LDOS (middle row, panels (b), (e), (h)), and the LMDOS (bottom row, panels (c), (f), (i)) for a range of values of $\abs{\vb{x}}/\xi$ in $D = 3$, and $\omega/\Delta$. Each column corresponds to a different value of the dimensionless impurity strength parameter, $J_{m} = 0$ (left column, panels (a), (b), (c)), $J_{m} = 0.5$ (middle column, panels (d), (e), (f)), and $J_{m} = 1.0$ (right column, panels (g), (h), (i)). The SC coherence length is $\xi/\lambda_{\rm F} = 3$, and  $f_{0}$ in \req{eq:ScalingFn} is $f_{0} = 0.27$.}
\end{figure}

In Fig.~\ref{fig:LMDOS}, we display the spin-up LDOS (top row, panels (a), (d), (g)), the spin-down LDOS (middle row, panels (b), (e), (h)) and the LMDOS (bottom row, panels (c), (f), (i)) as function of the frequency $\omega/\Delta$ and the position $\abs{\vb{x}}/\xi$ away from the impurity.

For illustration purposes, we take a short SC coherence length $\xi/\lambda_{\mathrm{F}} = 3$, which implies a value for the SC OP in units of the Fermi energy ($\varepsilon_{\mathrm{F}}$), $\Delta/\varepsilon_{\mathrm{F}} = \lambda_{\mathrm{F}}/\qty(2 \, \pi \, \xi) = 0.053$. Also, we evaluate the Green's function at $\omega\to\omega+i\eta$ where $\eta/\Delta = 0.0005$ in \reqs{eq:Spectra} to avoid divergencies. Each column in Fig. \ref{fig:LMDOS} corresponds to a different value of the dimensionless impurity strength $J_{m}$.
The energy $\varepsilon_{0}$ of the YSR bound state, \req{eq:YSRenergy}, changes sign when $J_{m} \rightarrow 1/J_{m}$. This has the effect of switching the sub-gap peaks in any of the Figs.~\ref{fig:LMDOS}-\ref{fig:OffSiteOSO}
, including their sign, as was explicitly verified by comparing the middle column panels with analogous plots for $J_{m} = 2.0$.

For a clean system, $J_{m} = 0$ (left column), there is no spin polarization and the LDOS is homogeneous with BCS coherence peaks at $\pm \abs{\Delta}$. As the impurity strength is increased (middle column), two sub-gap SP peaks appear at $\pm\varepsilon_0$: YSR bound states. Below the critical impurity strength ($J_m<1$), the state at $-\varepsilon_{0}$ has the same spin polarization as the impurity, while the one at $\varepsilon_0$ has the opposite polarization. The spin polarization of these peaks does not oscillate with position. At the same time, the above-gap continuum shows a small spin polarization that has the same direction as the impurity for both positive and negative frequencies, but oscillates with position according to the oscillations of the basis functions (Fig.~\ref{fig:Basisd=3} of Appendix~\ref{app:BasisFns}).

At critical impurity strength, $J_{m} = 1$ (right column), the two sub-gap peaks merge at zero-frequency. Remarkably, they do not cancel, but, instead display oscillation with position. As it will be shown in the following, similar features are also observed in the spin-triplet odd-$\omega$ local pair correlation. Finally, for $J_m>1$, the YSR states cross and their polarization is reversed.

\subsection{\label{sec:ResLocalPairCorr}Local pair correlations spectra}
\begin{figure}[!h]
\includegraphics[width=\scale \linewidth]{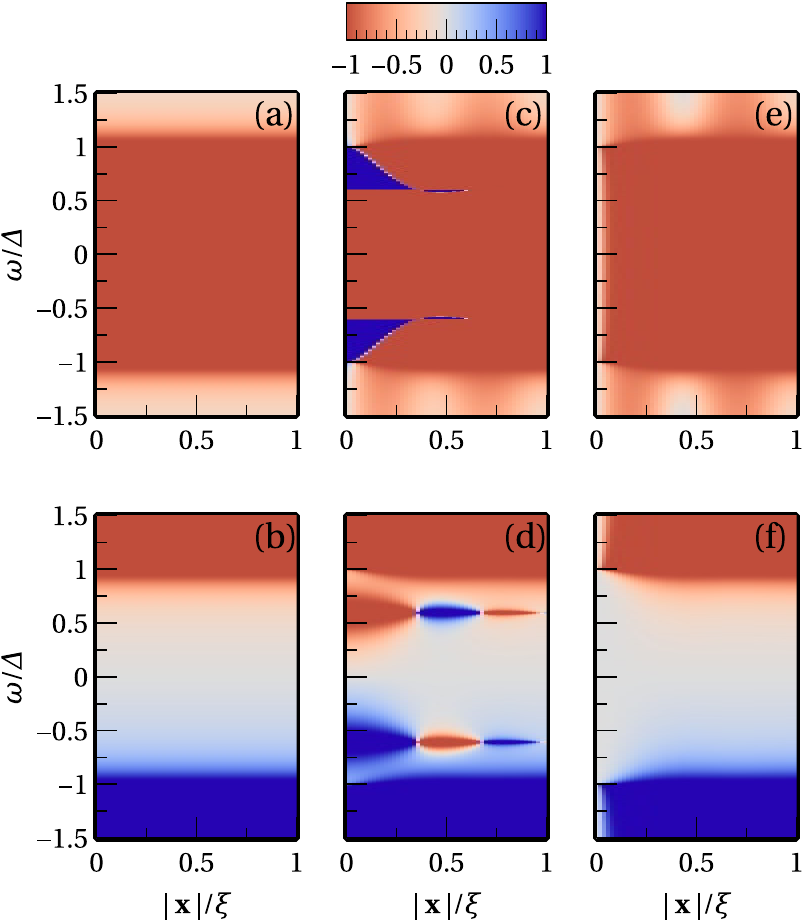}
\caption{\label{fig:OnSiteESE}Density plots for the real (panels (a), (c), and (e) in the top row) and imaginary part (panels (b), (d), and (f) in the bottom row) of $-\frac{1}{\pi \, N_{\rm F}} \, G^{e h}_{\qty(-,+,+)}\qty(\vb{x}, \vb{x}, \omega + \ii \, \eta)$ for a range of values of $\abs{\vb{x}}/\xi$, and $\omega/\Delta$. The choice of parameters is the same as in Fig.~\ref{fig:LMDOS}, and  $f_{0}$ in \req{eq:ScalingFn} is $f_{0} = 0.60$.}
\end{figure}

\begin{figure}[!h]
\includegraphics[width=\scale \linewidth]{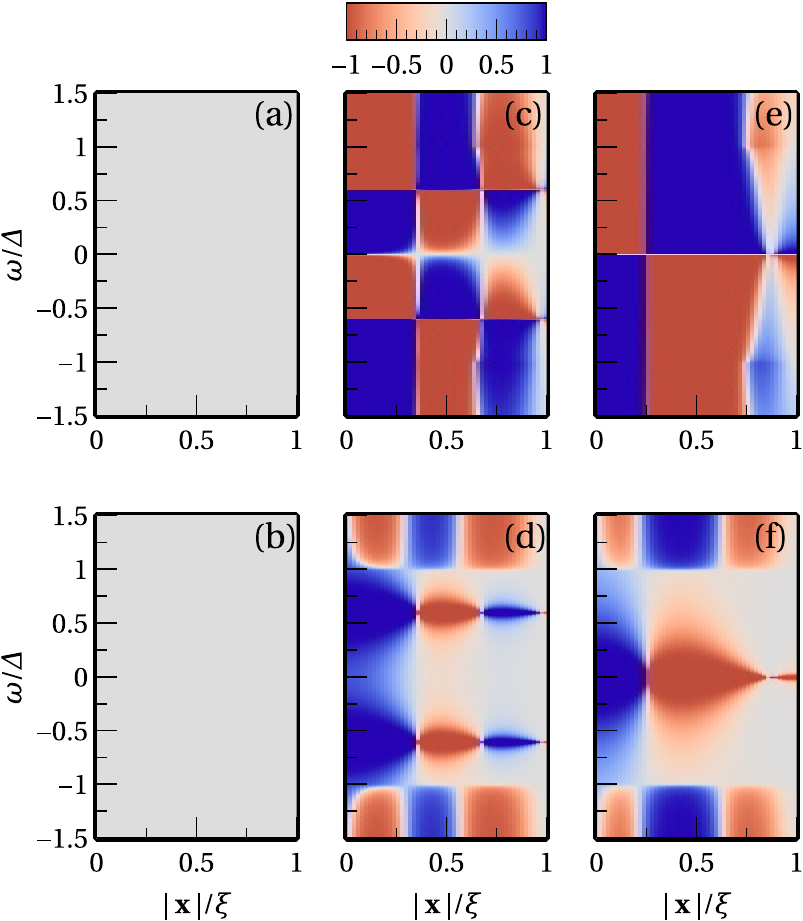}
\caption{\label{fig:OnSiteOTE}Density plots for the real (panels (a), (c), and (e) in the top row) and imaginary part (panels (b), (d), and (f) in the bottom row) of $-\frac{1}{\pi \, N_{\rm F}} \, G^{e h}_{\qty(+,+,-)}\qty(\vb{x}, \vb{x}, \omega + \ii \, \eta)$ for a range of values of $\abs{\vb{x}}/\xi$, and $\omega/\Delta$. The choice of parameters is the same as in Fig.~\ref{fig:LMDOS}, and  $f_{0}$ in \req{eq:ScalingFn} is $f_{0} = 0.11$.}
\end{figure}

Guided by the pair correlations of the conventional SC OP, which are on-site, we first consider local ($\vb{x} = \vb{x}'$) pair correlations in the vicinity of the impurity. For this choice, the symmetry under exchange of coordinates is necessarily even $P^{\ast} = +1$, so there are only $2$ possible components consistent with the $SP^{\ast}T^{\ast} = -1$ rule, namely even-$\omega$ spin-singlet $G^{e h}_{\qty(-, +, +)}\qty(\vb{x}, \vb{x}, \omega)$ and odd-$\omega$ spin-triplet $G^{e h}_{\qty(+, +, -)}\qty(\vb{x}, \vb{x}, \omega)$.

The conventional correlation belongs to the contribution $G^{e h}_{\qty(-,+,+)}$, displayed in Fig.~\ref{fig:OnSiteESE}. In the clean limit (panels (a) and (b)), the system is uniform and there is no $\vb{x}$ dependence. The correlation is real and even-$\omega$ for sub-gap frequencies ($-\abs{\Delta} < \omega < \abs{\Delta}$), while it becomes imaginary and odd for above-gap frequencies, as expected from a conventional superconductor. As the impurity strength is increased from $J_m=0$ (panels (c), (d), (e), and (f)), the features above the bulk gap persist. At intermediate impurity strengths, the build-up of YSR states at $\pm\varepsilon_{0}$ (for $J_{m} = 0.5$ $\varepsilon_{0} = 0.6 \, \abs{\Delta}$, in panels (c) and (d)) leads to additional features inside the superconducting gap. The real part of the $(-,+,+)$ component changes sign on both sides of the peak while the imaginary part has the same sign. This feature persists for the real and the imaginary part of any correlation function around the YSR, Figs.~\ref{fig:OnSiteOTE}-\ref{fig:OffSiteOSO}. Further we note that the parity of the real and imaginary parts of any correlation are opposite.~\cite{Bzudshek2015} The real part, being even in frequency, has the same sign between the YSR bound state energies ($\abs{\omega} < \varepsilon_{0}$) and changes sign as the frequency crosses the YSR bound state energies, vanishing at the gap edges ($\varepsilon_{0} < \abs{\omega} < \Delta$). The imaginary part, being odd in frequency, has sharp peaks around the YSR bound-state energies with opposite sign and spatial modulation determined by that of Fig.~\ref{fig:Basisd=3}. As the strength of the impurity approaches the critical value, $J_{m} = 1$, the two peaks with opposite sign merge at zero energy and cancel out. Therefore, the conventional superconducting correlations, $\qty(-, +, +)$, are suppressed for the sub-gap frequency region and in the proximity to the impurity. For $J_m>1$ the sub-gap features appear again, showing the same symmetry as in the $J_m=0.5$ case, with a global sign in the imaginary part. 

The odd-$\omega$ on-site component $G^{e h}_{\qty(+,+,-)}$ is displayed in Fig.~\ref{fig:OnSiteOTE}. Here we point out only the differences with Fig.~\ref{fig:OnSiteESE}. We note that the correlation is induced by the effect of the impurity (only from $\qty[G^{t}_{\rm imp}]^{e h}$) and, therefore, completely vanishes for a clean superconductor $(J_{m} = 0)$. At finite $J_m$, we observe the appearance of signatures outside and inside the gap.  This component is triplet with spin polarization in the direction of the impurity, as illustrated by Eq. (\ref{eq:ImpGreen3}). As shown in Fig.~\ref{fig:OnSiteOTE}, the real part is odd (and the imaginary part is even) in $\omega$. Differently from the previously commented case, as the peaks merge (panel Fig.~\ref{fig:OnSiteOTE}(f)), they do not cancel, but, enhance instead. The same enhancement persists also in the real part. The spatial modulation is identical to the one of Fig.~\ref{fig:OnSiteESE}, which is in agreement with the reciprocity of the two matrices having the same spatial dependence. One final feature is the small contribution  of the continuum states exhibiting the same spatial modulation than the YSR.

It is important to note that at the critical impurity strength the conventional $(-, +, +)$ pair correlation is completely suppressed in the vicinity of the impurity, as shown in Fig. ~\ref{fig:OnSiteESE} (e) and (f). Therefore only local component that survives is the $(+, +, -)$, which corresponds to the Berezinskii pairing, \ref{fig:OnSiteOTE} (panels (e), and (f)).

\subsection{\label{sec:ResNonLocalPairCorr}Non-local pair correlations spectra}
\begin{figure}[!h]
\includegraphics[width=\scale \linewidth]{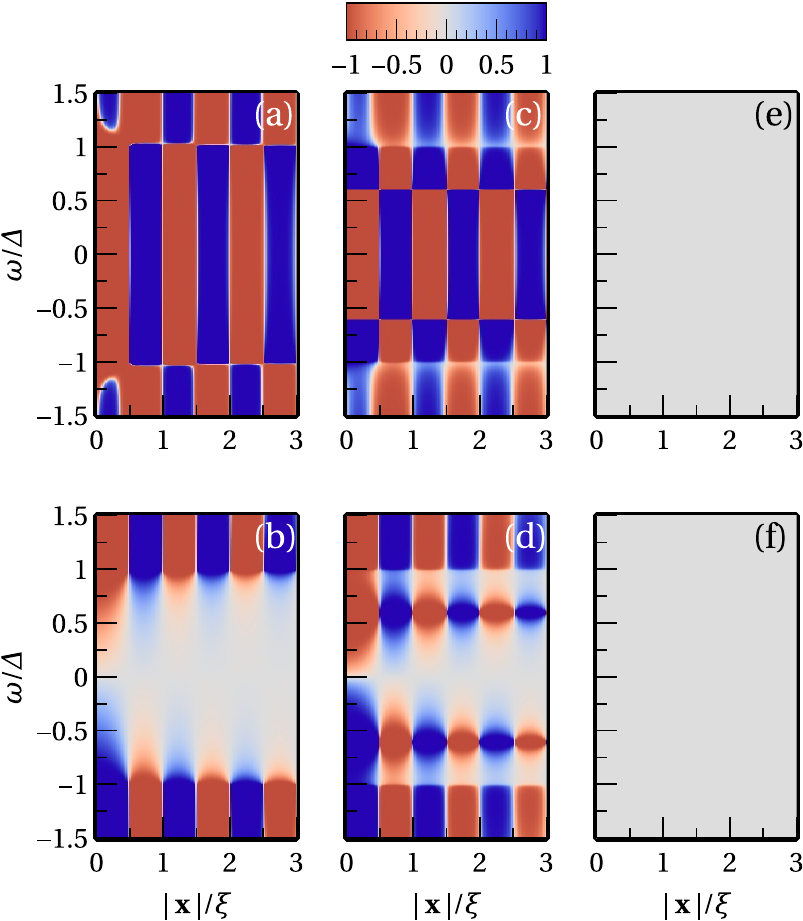}
\caption{\label{fig:OffSiteESE}Density plots for the real (panels (a), (c), and (e) in the top row) and imaginary part (panels (b), (d), and (f) in the bottom row) of $-\frac{1}{\pi \, N_{\rm F}} \, G^{e h}_{\qty(-,+, +)}\qty(\vb{x}, \vb{0}, \omega + \ii \, \eta)$ for a range of values of $\abs{\vb{x}}/\xi$, and $\omega/\Delta$. The choice of parameters is the same as in Fig.~\ref{fig:LMDOS}, and  $f_{0}$ in \req{eq:ScalingFn} is $f_{0} = 0.071$.}
\end{figure}

\begin{figure}[!h]
\includegraphics[width=\scale \linewidth]{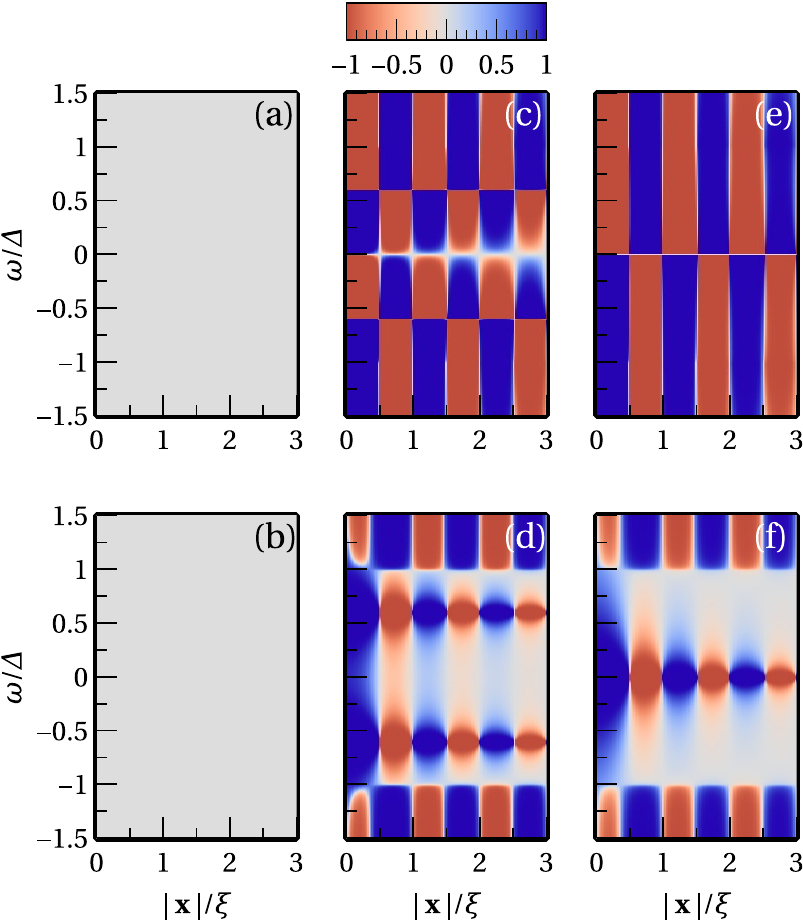}
\caption{\label{fig:OffSiteOTE}Density plots for the real (panels (a), (c), and (e) in the top row) and imaginary part (panels (b), (d), and (f) in the bottom row) of $-\frac{1}{\pi \, N_{\rm F}} \, G^{e h}_{\qty(+, +, -)}\qty(\vb{x}, \vb{0}, \omega + \ii \, \eta)$ for a range of values of $\abs{\vb{x}}/\xi$, and $\omega/\Delta$. The choice of parameters is the same as in Fig.~\ref{fig:LMDOS}, and  $f_{0}$ in \req{eq:ScalingFn} is $f_{0} = 0.064$.}
\end{figure}
\begin{figure}[!h]
\includegraphics[width=\scale \linewidth]{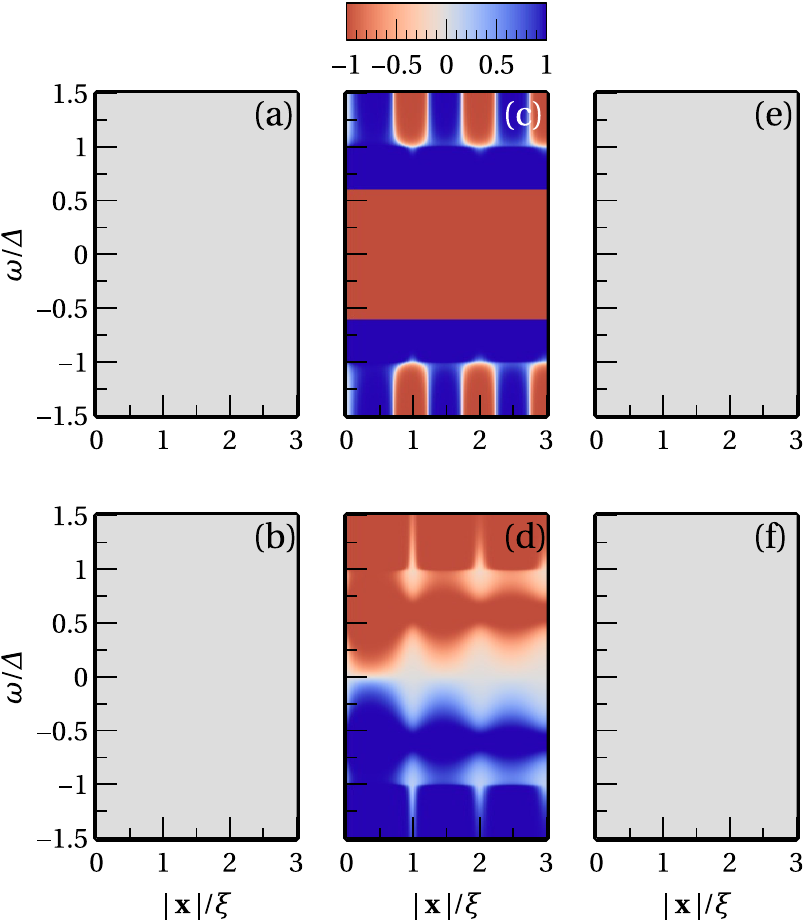}
\caption{\label{fig:OffSiteETO}Density plots for the real (panels (a), (c), and (e) in the top row) and imaginary part (panels (b), (d), and (f) in the bottom row) of $-\frac{1}{\pi \, N_{\rm F}} \, G^{e h}_{\qty(+, -, +)}\qty(\vb{x}, \vb{0}, \omega + \ii \, \eta)$ for a range of values of $\abs{\vb{x}}/\xi$, and $\omega/\Delta$. The choice of parameters is the same as in Fig.~\ref{fig:LMDOS}, and  $f_{0}$ in \req{eq:ScalingFn} is
$f_{0} = 0.020$.}
\end{figure}
\begin{figure}[!h]
\includegraphics[width=\scale \linewidth]{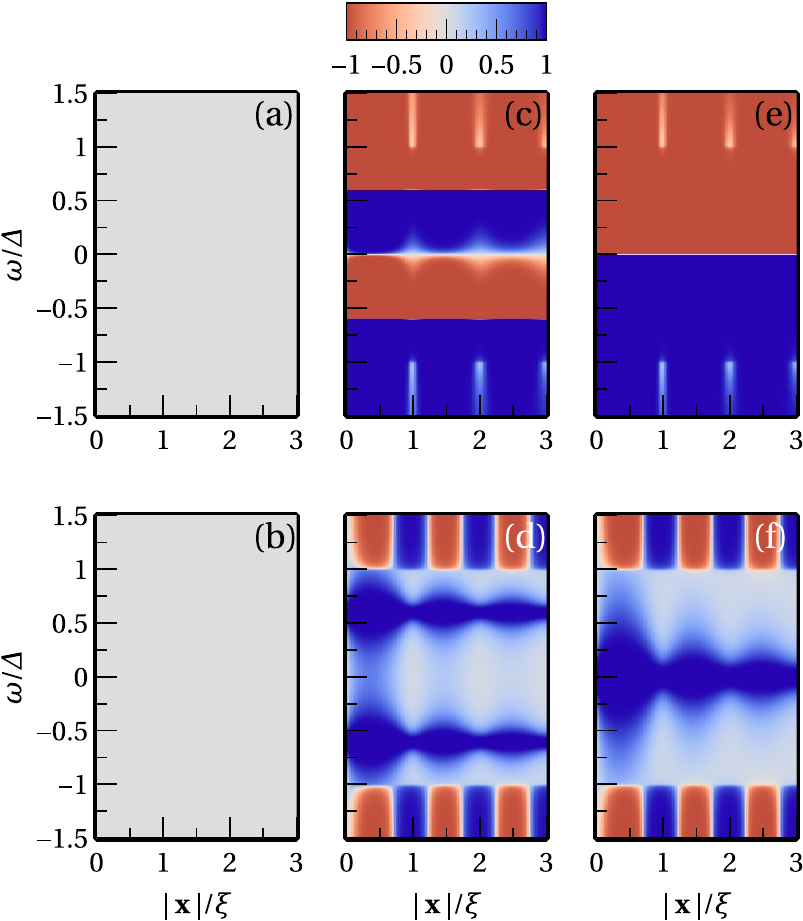}
\caption{\label{fig:OffSiteOSO}Density plots for the real (panels (a), (c), and (e) in the top row) and imaginary part (panels (b), (d), and (f) in the bottom row) of $-\frac{1}{\pi \, N_{\rm F}} \, G^{e h}_{\qty(-, -, -)}\qty(\vb{x}, \vb{0}, \omega + \ii \, \eta)$ for a range of values of $\abs{\vb{x}}/\xi$, and $\omega/\Delta$. The choice of parameters is the same as in Fig.~\ref{fig:LMDOS}, and  $f_{0}$ in \req{eq:ScalingFn} is 
$f_{0} = 0.062$.}
\end{figure}

Another possibility to consider is the non-local pair correlations. Because the impurity site is preferential, we are interested in fixing one of the coordinates at the impurity site $\vb{x}' = \vb{0}$. In Figs.~\ref{fig:OffSiteESE}-\ref{fig:OffSiteOSO},  we show the electron-hole separation, when a hole is created at the impurity site and an electron is annihilated at a site $\vb{x}$. 

Fig.~\ref{fig:OffSiteESE} displays the pair correlation with the same symmetry as the one in Fig.~\ref{fig:OnSiteESE}. In the absence of an impurity (panels (a) and (b)), it displays a characteristic periodicity with a period $\xi$. The spatial modulation is according to the different behavior of the basis functions for this choice of coordinates (Fig.~\ref{fig:Basis2d=3}). Unlike the local correlations, where the imaginary part is zero for sub-gap frequencies, in this case there is a finite component that decays with separation over a length scale determined by $1/\Re\qty(\kappa\qty(\omega))$, introduced in relation to \reqs{eq:GreenClean2}. As expected, the frequency dependence, including the parity and the behavior around the YSR bound state energies is the same as in the local case, since it is determined by the basis functions. Thus, at critical impurity strength, $J_m=1$, (Fig.~\ref{fig:OffSiteESE}(e), (f)) the imaginary and real parts vanish in accordance with the exact cancellation of the local correlation at the impurity site.

An analogous comparison may be made for the non-local (Fig.~\ref{fig:OffSiteOTE}) and local (Fig.~\ref{fig:OnSiteOTE}) correlations with symmetry $(+,+,-)$. It exhibits the same spatial dependence as Fig.~\ref{fig:OffSiteESE}. As in Fig.~\ref{fig:OnSiteOTE}, the real part of the correlation is odd in frequency while the imaginary one is even. Similarly to the local case, the $(+,+,-)$ correlation does not vanish for $J_m=1$, but increased instead. It means that, close to the critical impurity strength, the odd correlations dominate.

In addition to the two components discussed before, there are two more correlation components appearing in the non-local case. According to Table~\ref{tab:TableGeh}, they correspond to the case where the correlation is odd under exchange of spatial coordinates ($P^{\ast} = -1$). They are displayed in Figs.~\ref{fig:OffSiteETO} and \ref{fig:OffSiteOSO}, respectively.

As shown, both of the odd-parity non-local pair correlations require a non-zero impurity scattering, as illustrated by the vanishing of  Fig.~\ref{fig:OffSiteETO}(a), (b) and Fig.~\ref{fig:OffSiteOSO}(a), (b). For $J_m\neq1$, both components exhibit a finite value, where the symmetry of the real part is set by the $T^{\ast}$ value. For $J_m=1$, the (+,-,+) component vanishes while the (-,-,-), odd-$\omega$ gets enhanced.
By comparing the behavior at critical impurity strength on Fig.~\ref{fig:OffSiteETO}(e), (f), and Fig.~\ref{fig:OffSiteESE}(e), (f), we conclude that there exist another type of reciprocity. In this case the parity and spin symmetry are exchanged, while keeping the frequency symmetry unchanged, as hinted in Sec.~\ref{sec:Reciprocity}. We note that the the odd-parity pair correlations have spatial dependence determined by the only odd basis function $\psi_{4}$, which has a different spatial behavior than the even ones (Fig.~\ref{fig:Basis2d=3}), with no change in sign in both the real and imaginary part for sub-gap frequencies.

\section{\label{sec:Concl}Conclusions}
To conclude, we have demonstrated that the Yu-Shiba-Rusinov (YSR) bound states formed at the vicinity of magnetic impurities in conventional superconductors harbor spin-triplet, odd-$\omega$ pair correlations, which may be referred to as Berezinskii-YSR (BYSR) states. In case of spin-rotation invariant bulk SC, the spin polarization of the BYSR state is collinear with the impurity spin. We have demonstrated that the odd-frequency pairing dominates close to the impurity at the critical coupling strength, where YSR states merge at the middle of the gap. This leads to a regime where odd-frequency correlations can be probed experimentally. We note that the local odd-frequency component exhibits the same frequency and spatial dependence than the local magnetization density of states, which can be measured using  spin-polarized scanning tunneling spectroscopy. By solving exactly for the Green's function, we were able to identify a complete reciprocity as a function of the impurity coupling strength  under exchange of the $S$ and $T^{\ast}$ symmetries and with the same  parity coordinate, $P^{\ast}$. Finally, we have also demonstrated the presence of all the allowed non-local pair correlations consistent with the $SP^{\ast}OT^{\ast}=-1$ rule and summarized in Table \ref{tab:TableGeh}, showing that even-parity correlations are suppressed at the critical impurity strength.

\acknowledgements
The author thank Alexander Khajetoorians, and Johan Mentink for useful discussions. DK and AVB acknowledge funding  by VILLUM FONDEN via the Centre of Excellence for Dirac Materials (Grant No.  11744), Knut and Alice Wallenberg Foundation (Grant No. KAW 2018.0104), and the European Research Council ERC-2018-SyG HERO. RSS acknowledges funding from QuantERA project 2D hybrid materials as a platform for topological quantum computing.

\appendix

\section{\label{app:RealSpaceGcl}Real space clean superconductor Green's function in the wide-band approximation}
The real space Green's function is obtained by inverting \req{eq:BdGEqn1} and performing the momentum integral according to \req{eq:GreenClean1}. When performing the momentum integral the following steps are taken:
\begin{subequations}
\label{eq:MomentumIntegralSteps}
\begin{eqnarray}
& \frac{1}{L^{D}} \, \sum_{\vb{k}} \ee^{\ii \, k \, r \cos \theta} \, \hat{G}_{\rm cl}\qty(\xi_{k}; \omega) \nonumber \\
& = \frac{\Omega_{D}}{\qty(2 \pi)^{D}} \int \frac{\dd{\solid}}{\Omega_{D}} \int_{0}^{\infty} \dd{k} k^{D - 1} \ee^{\ii \, k \, r \cos \theta} \, \hat{G}_{\rm cl}\qty(\xi_{k}; \omega) \label{eq:MomentumIntegralStep1} \\
& \approx \frac{\Omega_{D} k^{D -1}_{\rm F}}{\qty(2 \pi)^{D} \, v_{\rm F}} \, \int \frac{\dd{\solid}}{\Omega_{D}} \, \ee^{\ii k_{\rm F} \, r \, \cos \theta} \nonumber \\
& \times \int_{-k_{\rm F} \, v_{\rm F}}^{\infty} \dd{\xi} \qty(1 + \frac{\xi}{k_{\rm F} \, v_{\rm F}})^{D - 1} \, \ee^{\ii \, \frac{r \, \cos \theta}{v_{\rm F}} \xi} \, \hat{G}_{\rm cl}\qty(\xi; \omega) \label{eq:MomentumIntegralStep2} \\
& \approx N_{\rm F} \, \int \frac{\dd{\solid}}{\Omega_{D}} \, \ee^{\ii k_{\rm F} \, r \, \cos \theta} \, \int_{-\infty}^{\infty} \dd{\xi} \ee^{\ii \, \frac{r \, \cos \theta}{v_{\rm F}} \xi} \, \hat{G}_{\rm cl}\qty(\xi; \omega). \label{eq:MomentumIntegralStep3}
\end{eqnarray}
\end{subequations}
Here, $\dd{\solid}$ stands for integration over the directions $\hat{\vb{k}}$ with $\Omega_{D}$ being the solid angle in $D$ dimensions. The angle $\theta$ is the angle between $\vb{r}$ and $\vb{k}$.
In step \req{eq:MomentumIntegralStep1}, the summation over momenta was converted to an integration over a continuous variable not restricted to the First Brillouin Zone, using $D$-dimensional spherical coordinates.  This approximation is valid as long as the Fermi wavevector is far away from the Brillouin Zone boundaries. In step \ref{eq:MomentumIntegralStep2}, the linear approximation of the dispersion relation $\xi_{k} \approx v_{\rm F} \, \qty(\vb{k} - k_{\rm F})$ was employed. Here, the Fermi momentum $k_{\rm F}$ is defined as $\xi_{k_{\rm F}} = 0$ and the Fermi velocity is the derivative of the dispersion relation evaluated at $k_{\rm F}$, $v_{\rm F} = d \xi_{k}/d k \vert_{k = k_{F}}$. This approximation is valid as long as the relevant energy scales $\delta \varepsilon$ that enter in the function $\hat{G}_{\rm cl}$ (such as $\Delta$ and $\omega$), as well as in the exponential ($v_{\rm F}/r$) satisfy the constraint $\abs{\delta \varepsilon} \ll \abs{m^{\ast}} \, v^{2}_{\rm F}/2$. Here, $m^{\ast}$ is the band mass $\qty(m^{\ast})^{-1} = d^{2} \xi_{k}/d k^{2} \vert_{k = k_{\rm F}}$. For a parabolic band, the right-hand side of this inequality is simply the Fermi energy $\varepsilon_{\rm F}$, i.e. the energy difference between the bottom of the dispersion relation at $k = 0$ and the Fermi level at $k = k_{\rm F}$. For a linear band $\varepsilon_{\rm F} = v_{\rm F} \, k_{\rm F}$. So, the linear approximation is valid as long as the energy scales of the problem are much lower than the Fermi energy. In step \req{eq:MomentumIntegralStep3}, in accordance with the above assumptions, the Fermi energy is taken to infinity. To make the integrals convergent, any high powers of $\xi$ were neglected as well, essentially approximating the DOS per spin component as being flat with energy, and equal to its value at the Fermi level:
\begin{equation}
\label{eq:NormDos}
N_{\rm F} = \frac{\Omega_{D} k^{D - 1}_{\rm F}}{\qty(2 \pi)^{D} \, v_{\rm F}} = \frac{\Omega_{D}}{\lambda^{D}_{\rm F} \, \varepsilon_{\rm F}},
\end{equation}
where $\lambda_{\rm F} = 2\pi/k_{\rm F}$ is the Fermi wavelength. This approximation is in agreement with the above quasi-continuum approximation as long as $k_{\rm F} \ll \pi/a$, with $a$ being the lattice constant. All these approximations are referred to as the wide-band approximation.

Inverting \req{eq:BdGEqn1}, we find
\begin{equation}
\label{eq:BdGEqn2}
\hat{G}_{\rm cl}\qty(\xi; \omega) = -\frac{1}{\xi^{2} + \abs{\Delta}^{2} - \omega^{2}} \, \begin{pmatrix}
\omega + \xi & \Delta \\
\Delta^{\ast} & \omega - \xi
\end{pmatrix},
\end{equation}
where the integrals over $\xi$ in \req{eq:MomentumIntegralSteps} are performed using the following expressions:
\begin{subequations}
\label{eq:XiIntExprs}
\begin{eqnarray}
& \int_{-\infty}^{\infty} \dd{\xi} \frac{\ee^{\ii \, a \, \xi}}{\xi^{2} + z} = \frac{\pi}{\sqrt{z}} \, \ee^{-\abs{a} \, \sqrt{z}}, \label{eq:XiIntExpr1} \\
& \int_{-\infty}^{\infty} \dd{\xi} \frac{\xi \, \ee^{\ii \, a \, \xi}}{\xi^{2} + z} = \ii \, \pi \, {\rm sgn}\qty(a) \, \ee^{-\abs{a} \, \sqrt{z}}. \label{eq:XiIntExpr2}
\end{eqnarray}
\end{subequations}
These integrals are valid for real $a$ and $\Re\qty[\sqrt{z}] > 0$. In \req{eq:XiIntExpr2}, it is assumed that ${\rm sgn}(0) = 0$, in accordance with the Cauchy principal value of the integral. Comparing with \req{eq:BdGEqn2}, one can see that $a = r \, \cos \theta/v_{\rm F}$, and $z = \abs{\Delta}^{2} - \omega^{2}$.

\begin{table*}[!th]
\begin{ruledtabular}
\begin{tabular}{lccc}
$D$ & $I_{D}\qty(z)$ & $\qty(\phi_{c})_{D}\qty(r, \omega)$ & $\qty(\phi_{s})_{D}\qty(r, \omega)$ \\
\hline
$1$ & $\ee^{-z}$ & $\ee^{-\kappa\qty(\omega) \, r} \, \cos\qty(k_{\rm F} \, r)$ & $-\ee^{-\kappa\qty(\omega) \, r} \, \sin\qty(k_{\rm F} \, r)$\\
$3$ & $\frac{1 - \ee^{-z}}{z}$ & $\frac{-\kappa\qty(\omega) - \ee^{-\kappa\qty(\omega) \, r} \, \qty[\kappa\qty(\omega) \, \cos\qty(k_{\rm F} \, r) - k_{\rm F} \, \sin\qty(k_{\rm F} \, r)]}{\qty[k^{2}_{\rm F} + \kappa^{2}\qty(\omega)] \, r}$ & $\frac{-k_{\rm F} + \ee^{-\kappa\qty(\omega) \, r} \, \qty[k_{\rm F} \, \cos\qty(k_{\rm F} \, r) + \kappa\qty(\omega) \, \sin\qty(k_{\rm F} \, r)]}{\qty[k^{2}_{\rm F} + \kappa^{2}\qty(\omega)] \, r}$
\end{tabular}
\end{ruledtabular}
\caption{\label{tab:TableFuncs}Analytic expressions for the functions defined in \reqs{eq:AngleInt} in $D = 1$ and $D = 3$ that determine the spatial dependence of the Green's function for the clean system.  For $\omega < \abs{\Delta}$, $\kappa\qty(\omega)$ is real (and positive), which gives exponential decay in $D = 1$. In $D = 3$, besides the exponentially decaying part there is a constant part in the numerator. For $\omega > \abs{\Delta}$, $\kappa$ is imaginary which shifts the oscillation period in $D = 1$ and in $D = 3$. In both cases the overall decay is as $1/r$ from the $r$ in the denominator.}
\end{table*}

The angular integrals have the following expressions
\begin{subequations}
\label{eq:AngleInt}
\begin{eqnarray}
& \qty(\phi_{c})_{D}\qty(r, \omega) = \int \frac{\dd{\solid}}{\Omega_{D}} \, \ee^{-r \, \qty(\kappa\qty(\omega) \, \abs{\cos \theta} - \ii \, k_{\rm F} \, \cos \theta)} \nonumber \\
& = \frac{1}{2} \, \left\lbrace I_{D}\qty[\qty(\kappa\qty(\omega) + \ii \, k_{\rm F}) \, r] \right. \nonumber \\
& \left. + I_{D}\qty[\qty(\kappa\qty(\omega) - \ii \, k_{\rm F}) \, r] \right\rbrace, \label{eq:AngIntc} \\
& \qty(\phi_{s})_{D}\qty(r, \omega) = \ii \, \int \frac{\dd{\solid}}{\Omega_{D}} \, \ee^{-r \, \qty(\kappa\qty(\omega) \, \abs{\cos \theta} - \ii \, k_{\rm F} \, \cos \theta)} \, {\rm sgn}\qty(\cos \theta) \nonumber \\
& = \frac{1}{2 \, \ii} \, \left\lbrace I_{D}\qty[\qty(\kappa\qty(\omega) + \ii \, k_{\rm F}) \, r] \right. \nonumber \\
& \left. - I_{D}\qty[\qty(\kappa\qty(\omega) - \ii \, k_{\rm F}) \, r] \right\rbrace, \label{eq:AngInts} \\
& I_{D}\qty(z) = \int \dd{W_{\hat{\vb{k}}}} \ee^{-z \, \cos \theta} \nonumber \\
& = \Gamma\qty(\frac{D}{2}) \, \qty(\frac{z}{2})^{-1 + \frac{D}{2}} \qty[I_{D/2 - 1}\qty(z) - \mathbf{L}_{D/2 - 1}\qty(z)], \label{eq:SpecFunc}
\end{eqnarray}
\end{subequations}
where $\dd{W_{\hat{\vb{k}}}} = \frac{\dd{\solid} \, \Theta\qty(\cos\theta)}{\int \dd{\solid} \, \Theta\qty(\cos \theta)} $ is a normalized integration measured over only the hyperhemisphere where $\cos \theta > 0$, $\kappa\qty(\omega) = \rad/v_{\rm F}$, $I_{\nu}\qty(z)$ is the modified Bessel function of the first kind of order $\nu$ and $\mathbf{L}_{\nu}\qty(z)$ is the modified Struve function of order $\nu$. For $D = 1$ and $D = 3$, the special function expressions that appear in \reqs{eq:AngleInt} have an elementary form listed in Table~\ref{tab:TableFuncs}. Using thees results, we obtain \req{eq:GreenClean2}.

\section{\label{app:TmatExprs}$T$-matrix expression}
The Dyson equation \req{eq:DysonRealSpace} can be solved by employing a T-matrix approach. Namely, the following expression holds:
\begin{subequations}
\label{eq:TmatrixApproach}
\begin{eqnarray}
& \check{G}_{\rm imp}\doubargs = \check{G}_{\rm cl}\qty(\vb{x}, \vb{0}; \omega) \cdot \check{T}\qty(\omega) \cdot \check{G}_{\rm cl}\qty(\vb{0}, \vb{x}'; \omega), \label{eq:ImpGreen2} \\
& \check{T}^{-1}\qty(\omega) = \frac{1}{U_{m}} \, \check{P}^{-1} - \check{G}_{\rm cl}\qty(\vb{0}, \vb{0}; \omega). \label{eq:TmatrixEqn}
\end{eqnarray}
\end{subequations}
Having in mind \reqs{eq:ImpVertex} and \rref{eq:GreenCleanSemiClassNambu}, the solution of \req{eq:TmatrixEqn} for the T-matrix is given by
\begin{equation}
\label{eq:TmatrixNambuSpin}
\check{T}\qty(\omega) = \frac{U_{m}}{D_{\rm YSR}\qty(\omega)} \, \left[ \hat{N}^{s}_{\rm T}\qty(\omega) \otimes \hat{\sigma}_{0} + \hat{N}^{t}_{\rm T}\qty(\omega) \otimes \ns\right],
\end{equation}
where the denominator $D_{\rm YSR}\qty(\omega)$ is given by \req{eq:DYSR}. 

The superscript $s$ refers to the singlet component, and $t$ refers to the triplet component, given by the polarization axis $\vb{n}$ of the magnetic impurity atom. The numerators that enter in \req{eq:TmatrixNambuSpin}, $\hat{N}^{s}_{\rm T}\qty(\omega)$ and $\hat{N}^{t}_{\rm T}\qty(\omega)$, have the following structure in Nambu space:
\begin{widetext}
\begin{subequations}
\label{eq:TmatrixNumsNambu}
\begin{eqnarray}
& \hat{N}^{s}_{\rm T}\qty(\omega) = J_{m} \, \rad \, \left[\qty(1 + J^{2}_{m}) \, \omega \, \hat{\tau}_{0} + \qty(1 - J^{2}_{m}) \, \hat{\Delta} \right], \label{eq:TmatrixNsingl} \\
& \hat{N}^{t}_{\rm T}\qty(\omega) = \qty[ \qty(1 + J^{2}_{m}) \, \omega^{2} - \qty(1 - J^{2}_{m}) \, \abs{\Delta}^{2} ] \, \hat{\tau}_{0} - 2 J^{2}_{m} \, \omega \, \hat{\Delta}. \label{eq:TmatrixNtrip}
\end{eqnarray}
\end{subequations}
\end{widetext}
\begin{figure*}[!h]
\includegraphics[width= 0.75 \linewidth]{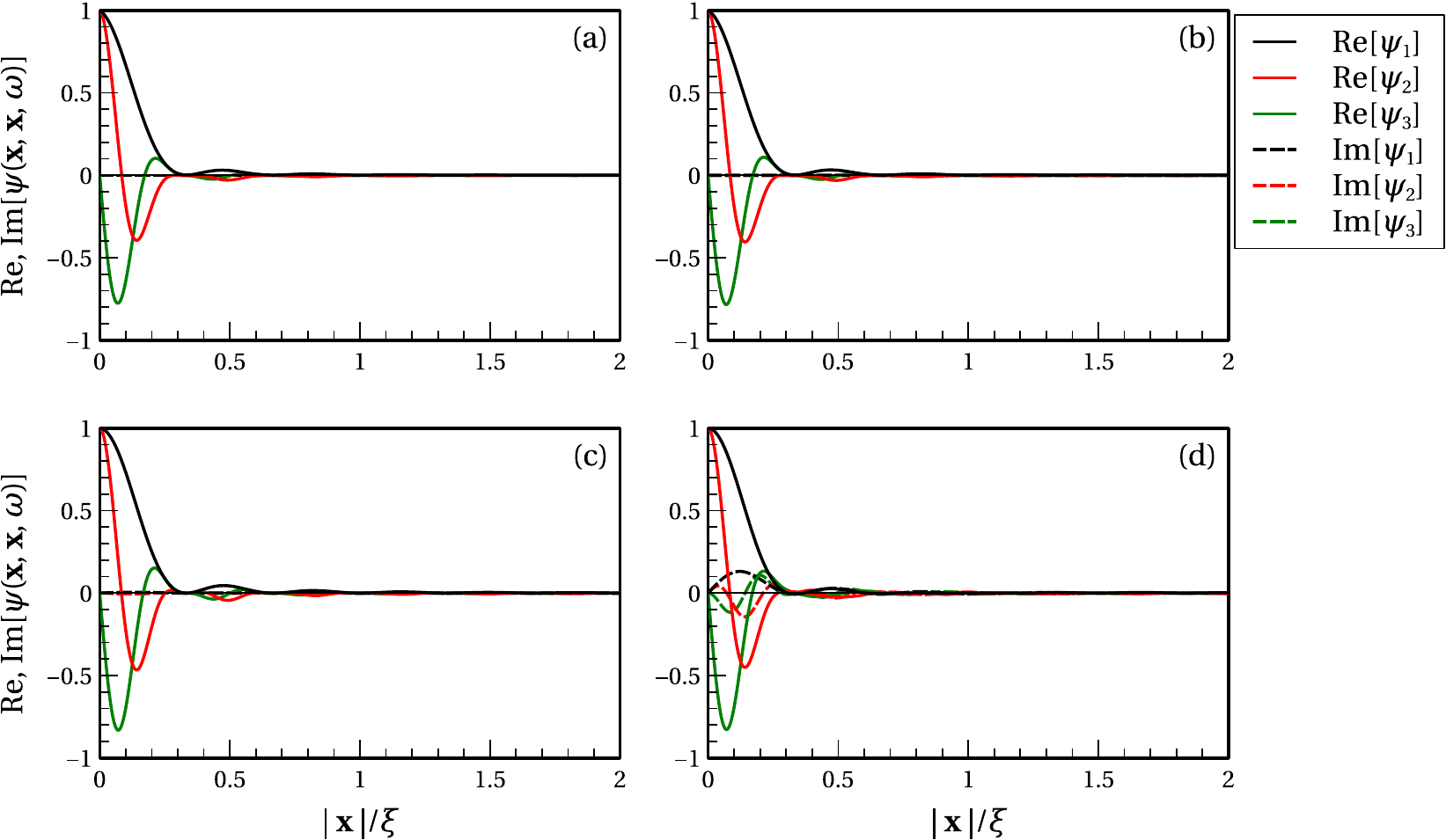}
\caption{\label{fig:Basisd=3}Real (solid lines) and imaginary (dashed lines) part of the retarded ($\omega \rightarrow \omega + \ii \, \eta$) basis functions $\psi_{a}$ \reqs{eq:BasisFns} as a function of $\abs{\vb{x}}/\xi$ for $\vb{x} = \vb{x}'$ and $D = 3$ dimensions. We represent $\psi_{1}$ (black lines), $\psi_{2}$ (red lines) and 
$\psi_{3}$ (green lines). We note $\psi_{4}=0$ as it is odd under position exchange. The SC OP is chosen so that $\xi/\lambda_{\rm F} = 3$. Different panels are fixed at different frequency: Panel $\omega/\Delta = 0$ (a), $\omega/\Delta = 0.5$ (b), $\omega/\Delta = 1.0$ (c), and $\omega/\Delta = 2.0$ (d).}
\end{figure*}

\section{\label{app:BasisFns}Spatial dependence of Basis functions}
\begin{figure*}[!h]
\includegraphics[width=0.75 \linewidth]{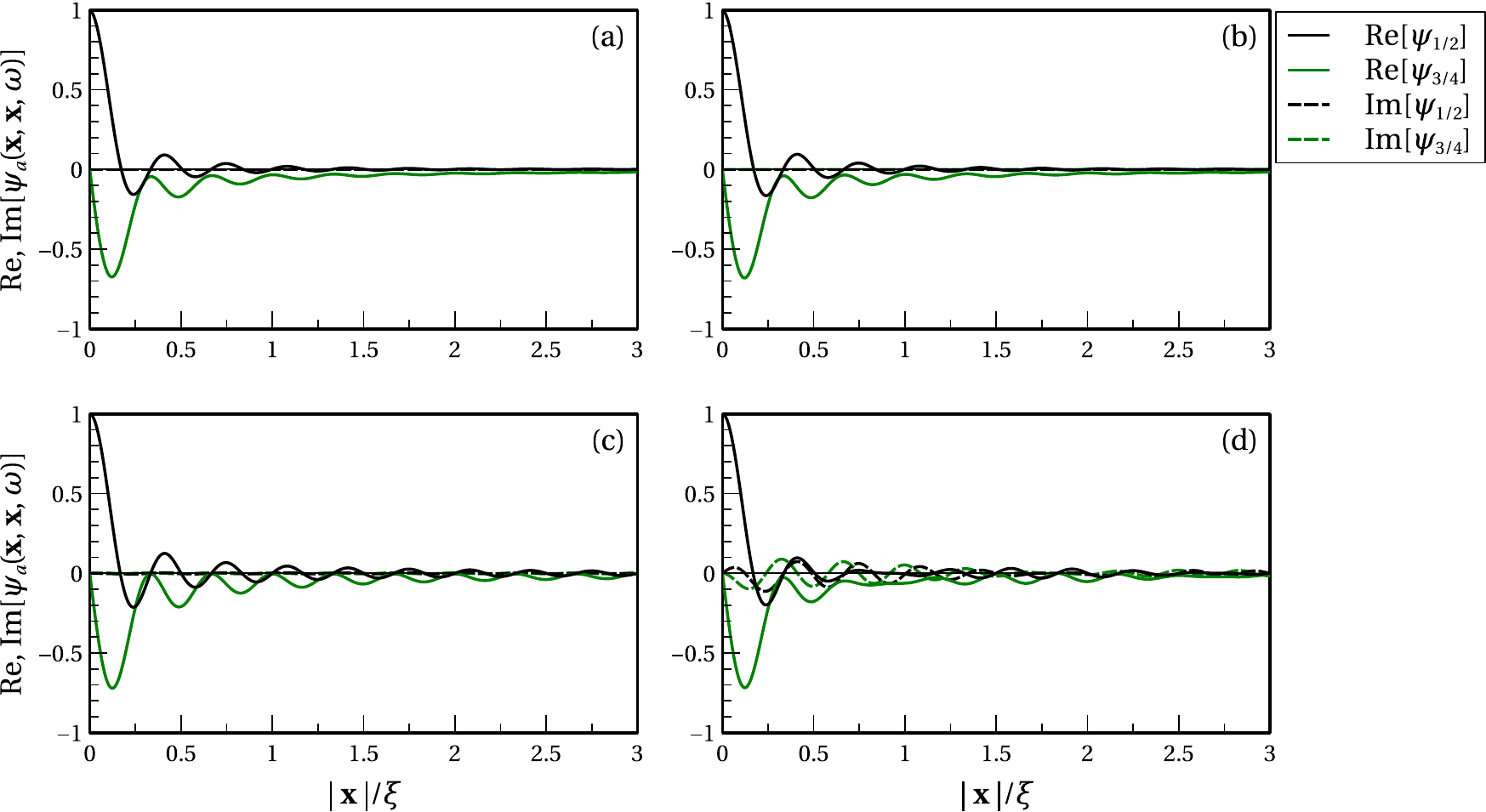}
\caption{\label{fig:Basis2d=3}Real (solid lines) and imaginary (dashed lines) part of the retarded ($\omega \rightarrow \omega + \ii \, \eta$) basis functions
$\psi_{a}$ \reqs{eq:BasisFns} as a function of $\abs{\vb{x}}/\xi$ for $\vb{x}' = \vb{0}$ and $D = 3$ dimensions. The black line represents the behavior for $a = 1$ and $a = 2$ while the green one represents the $a = 3$ and $a = 4$ functions. The SC OP is chosen so that $\xi/\lambda_{\rm F} = 3$. Different panels are fixed at different frequency: $\omega/\Delta = 0$ (a), $\omega/\Delta = 0.5$ (b), $\omega/\Delta = 1.0$ (c), and $\omega/\Delta = 2.0$ (d).}
\end{figure*}


The pair correlations in Sec.~\ref{sec:ResLocalPairCorr} and Sec.~\ref{sec:spLDOS} are evaluated at $\vb{x} = \vb{x}'$ in $D = 3$. In this section we analyze the spatial dependence of the basis functions \req{eq:BasisFns} for this choice of coordinates, for several values of frequency within and out of the SC gap $\abs{\Delta}$ and for the same choice of SC coherence length $\xi/\lambda_{\rm F} = 3$. Because $\psi_{4}$ is odd under the exchange $\vb{x} \leftrightarrow \vb{x}'$ by construction, it is identically zero. At $r = 0$, $\phi_{c} = 1$, and $\phi_{s} = 0$, so $\psi_{1} = \psi_{2} = 1$ and $\psi_{3} = 0$ for any $\omega$. All of these functions show oscillatory behavior with a characteristic period equal to $\lambda_{F}/\xi = 1/3$. In addition, $\psi_{2}$ and $\psi_{3}$ have an additional node at a value smaller than $\lambda_{\rm F}/\xi$. The imaginary part is zero for sub-gap frequencies and appears only for frequencies higher than the gap. We note there is an overall envelope determined by the squares of the functions $\qty(\phi_{c/s})_{3}(r; \omega)$ given in  Table~\ref{tab:TableFuncs}. 



Also, having in mind the non-local pair correlations presented in Sec.~\ref{sec:ResNonLocalPairCorr}, we plot in Fig.~\ref{fig:Basis2d=3} the basis functions \reqs{eq:BasisFns} for $\vb{x}' = \vb{0}$ as a function of $r = \vb{x}$ for the same choice of frequencies and parameters as in Fig.~\ref{fig:Basisd=3}. Similarly to the local case, the imaginary part vanishes for sub-gap frequencies and different from zero only above the superconducting gap. We note that the overall envelope decay is much slower, because these functions behave as the first powers of $\qty(\phi_{c/s})_{3}(r; \omega)$ given in Table~\ref{tab:TableFuncs}.

\bibliography{oddwMagImp_References}
\end{document}